# THz detection and amplification using plasmonic Field Effect Transistors driven by DC drain currents

Yuhui Zhang and Michael Shur[†]

*Department of Electrical, Computer, and Systems Engineering, Rensselaer Polytechnic Institute, Troy, New York 12180, USA*

[†] Corresponding author: Michael Shur (shurm@rpi.edu)

**Abstract**

We report on the numerical and theoretical results of sub-THz and THz detection by a current-driven InGaAs/GaAs plasmonic Field-Effect Transistor (TeraFET). New equations are developed to account for the channel length dependence of the drain voltage and saturation current. Numerical simulation results demonstrate that the effect of drain bias current on the source-to-drain response voltage ($dU$) varies with the device channel length. In a long-channel TeraFET where plasmonic oscillations cannot reach the drain, $dU$ is always positive and rises rapidly with increasing drain current. For a short device in which plasmonic oscillations reach the drain, the current-induced nonuniform electric field leads to a negative response, agreeing with previous observations. At negative $dU$, the amplitude of the small-signal voltage at the drain side becomes larger than that at the source side. Thus, the device effectively serves as a THz amplifier in this condition. Under the resonant mode, the negative response can be further amplified near the resonant peaks. A new expression of $dU$ is proposed to account for this resonant effect. Based on those expressions, a current-driven TeraFET spectrometer is proposed. The ease of implementation and simplified calibration procedures make it competitive or superior compared with other TeraFET-based spectrometers.

## 1. Introduction

The detection of sub-THz and THz signals by plasmonic THz detectors (TeraFETs) has been intensively studied in the last few decades [1-7]. Featuring high tunability, high sensitivity, large detection range, and fast response time, TeraFETs are promising for high-frequency applications such as THz sensing [8, 9], imaging [10-12], VLSI testing [13, 14], and beyond-5G communication [8, 15, 16]. The TeraFET applications, including the 300 GHz communication [15], rely heavily on detection sensitivity. Therefore, improving the detection responsivity of TeraFETs becomes a key topic in this area of study. Although having a remarkable theoretically predicted maximum responsivity [1], the existence of scatterings in the two-dimensional electron gas (2DEG) channel significantly impairs the device response [2, 17]. In addition, the non-ideal THz signal coupling and parasitics also lead to response degradation [18-20]. As a result, the observed voltage response in TeraFETs can be orders of magnitude lower than the analytical predictions.

To improve the TeraFET responsivity and enable its potential applications, several techniques have been explored, including the adoption of new materials [6, 21, 22] and non-uniform device structures [20, 23-25]. Using new materials in TeraFET systems is straightforward and proved to be effective. The choice of materials depends on the specific application scenarios. For example, p-Diamond was demonstrated to be a superior candidate for continuous-wave (CW) detection in the sub-THz band and short-channel devices [6, 22]. While for pulse detection, graphene-based TeraFETs might have better performance compared to other more traditional material systems [26]. Apart from the material consideration, the non-uniform asymmetrical TeraFETs have also been intensively studied. The non-uniformity in those devices can be realized by the multi-gate [20, 27], spatially varying gate capacitance [23, 24], non-uniform threshold voltage [23], and using the phase difference of the THz radiation signals coupled to the source and drain [28-31]. With these designs, the distributions of carrier density ($n$), static electric field ($E$), and plasma wave velocity ($s$) can be modified, thereby affecting the carrier transport and THz detection performance [24, 32]. In our recent work [23], we showed that a non-uniform 2DEG channel formed by the spatially varying gate capacitance or threshold voltage can effectively tune the THz response voltage of a given TeraFET. The driving force of the response modulation, in this case, originates from the formation of carrier density gradient and thus the modified dispersion relation in the 2DEG channel. In addition to the gate capacitance and threshold modulations, the multi-gate design was also applied. For example, in [20, 27, 33, 34] the multi-gates were concatenated to form a TeraFET dense array, in which the device terminals were split into fingers and nested together to form the repeated unit cells. Such a multi-finger structure enhanced the asymmetry and served as an effective antenna coupling incident THz signals. Consequently, the device's responsivity was strengthened.

Apart from the structural considerations, DC biasing is also an important factor affecting TeraFET performance [5, 35-38]. Ideally, the source of the TeraFET is grounded while the drain is left open

[1, 39]. A finite DC bias is applied to the gate controlling the 2DEG density underneath. When a direct current, instead of the open condition, is applied to the drain, the device could be unstable due to the plasmonic instability, causing periodic oscillation and THz generation near the drain side [36, 40-43]. This instability can be suppressed by the friction and viscosity effects in the 2DEG channel [17, 44]. The DC drain current also induces a DC electric field along the channel, and the DC gate-to-channel voltage $U_0$ becomes coordinate-dependent [5, 35]. As a result, the electron density turns non-uniform and alters the transport properties of the channel. As suggested in [5] and [35], the source-to-drain DC voltage response $dU$ rises exponentially with the increasing drain bias current as the device approaches the saturation region. The abrupt changes in the electric field near the drain were believed to be the main contributor to this response boom. Besides, the device channel length could significantly alter the detection performance of current-driven TeraFETs. A negative response was predicted for short-sample detections under relatively high current biases.

In this work, we use a one-dimensional (1D) hydrodynamic numerical model to simulate the sub-THz and THz response of a current-driven InGaAs/GaAs TeraFET. The simulation results provide insight into the device physics that allowed us to improve the analytical models and propose a new type of THz spectrometer based on the developed analytical expressions. The paper is organized as follows: in Section 2 the hydrodynamic model is briefly introduced. Theories and simulation results related to the DC operation of the current-driven TeraFET are discussed in Section 3. Section 4 describes the AC operation and THz detection performance. In Section 5, we propose a new TeraFET-based spectrometer based on the current-driven detection theory and discuss its pros and cons compared to similar devices proposed recently [28, 29, 45]. Finally, Section 6 presents concluding remarks.

**2. Hydrodynamic Model**

The numerical model used in this work is based on the classical hydrodynamic model for 2DEG systems [46-48]. The two main equations are the continuity equation and the Navier-Stocks equation, as shown below

$$\frac{\partial n}{\partial t} + \nabla \cdot (n\mathbf{v}) = 0 \tag{1}$$

$$\frac{\partial \mathbf{v}}{\partial t} + (\mathbf{v} \cdot \nabla)\mathbf{v} + \gamma \mathbf{v} + \frac{e}{m^*}\nabla U - \nu \nabla^2 \mathbf{v} = 0 \tag{2}$$

where $n$, $v$, and $m^*$ are the density, drift velocity, and effective mass of the electrons, respectively. $\gamma = 1/\tau = e/\mu m^*$ is the momentum relaxation rate, and $\mu$ is the drift mobility. $\upsilon$ is the kinematic viscosity of the electron fluid. $U = V_{gs} - V_{th} - V_{ch}$ is the gate-to-channel voltage, and $V_{gs}$, $V_{th}$, and $V_{ch}$ stand for the DC gate-to-source voltage, device threshold voltage, and channel voltage, respectively.

The gated 2DEG density under a given $U$ is determined by a unified charge control model (UCCM) [49, 50]:

$$n(U) = \frac{C_g \eta V_t}{e} \ln\left(1 + \exp\left(\frac{U}{\eta V_t}\right)\right), \quad V_t = \frac{k_B T}{e} \tag{3}$$

where $V_t = k_B T/e$ is the thermal voltage, $k_B$ is the Boltzmann constant, and $\eta$ is an ideality factor. Eq. (1)-(3) are the core of our numerical model, which does not account for the electron heating at a large drain bias. Our estimates show that accounting for the electron heating might affect the resonant detection regime (due to a commensurate decrease in the effective momentum relaxation time) [47, 51] but will not affect the qualitative results related to the non-resonant detection.

For a drain current-driven device, the terminal boundary conditions are given by [5, 36]

$$\begin{cases} U(0,t) = U_0(0) + V_{am} \cos \omega t \\ en(L,t)v(L,t) = J_d \end{cases} \tag{4}$$

where $U_0$ is the DC voltage swing. For a grounded source, $U_0(0) = U_g = V_g - V_{th}$, where $V_g$ is the DC gate bias. Note that we assumed a 1D geometry and $x = 0 \sim L$ represents the position from source to drain, where $L$ is the total channel length. $J_d$ is the DC drain current density, and a positive $J_d$ indicates a drain-to-source current flow. $V_{am}$ and $\omega = 2\pi f$ represent the amplitude and angular frequency of THz radiation, respectively.

The above equations and boundary conditions are discretized and solved using COMSOL software [52] by the finite-element method. More details on this model can be found in [26, 46, 53].

To validate this model, we compare the model simulation results with the relevant experimental data. The subject experimental data come from [35], in which an InGaAs/GaAs pseudomorphic high electron mobility transistor (p-HEMT) was used to detect a 1.63 THz signal under different biasing conditions. The parameter setup in our simulation is given in Table I, which follows the original experimental setup.

Table I. Parameters of InGaAs/GaAs heterostructure used in the simulation

| Param. | $\mu$ (cm$^2$/Vs) | $m^*$ ($m_0$) | $L$ (nm) | $V_{am}$ (mV) | $f$ (THz) | $T$ (K) | $v$ (cm$^2$/Vs) | $V_{th}$ (V) |
|---|---|---|---|---|---|---|---|---|
| Value | 3700 | 0.063 | 500 | 2 | 1.63 | 300 | 16.9 | 0.48 |

Fig. 1 shows the comparison of the experimental, analytical, and COMSOL simulation data under similar conditions. The analytical curves follow [5]

$$dU = \frac{V_{am}^2}{4\eta V_t [1+\exp(-\frac{U_g}{\eta V_t})] \ln[1+\exp(\frac{U_0(L)}{\eta V_t})]} \quad (5)$$

Here the ideality factor $\eta = 1.12$ [35]. Note that the values of $U_0(L)$ are taken from either the direct measurement (Fig. 1(a)) or the simulation (Fig. 1(b)), so that Eq. (5) can cover the subthreshold region. As seen from Fig. 1, for a given $V_{gs}$ the THz response rises rapidly with the increase of $I_d$. In the non-saturation region, the simulation data in Fig. 1(b) demonstrated a good agreement with both the experimental results in Fig. 1(a) and the analytical curves, indicating that our numerical model is valid in this region. After reaching the saturation condition, the slope of $dU$ changes due to the velocity saturation [54], as seen in Fig. 1(a). In this work, we only focus on the non-saturating operation of the device.

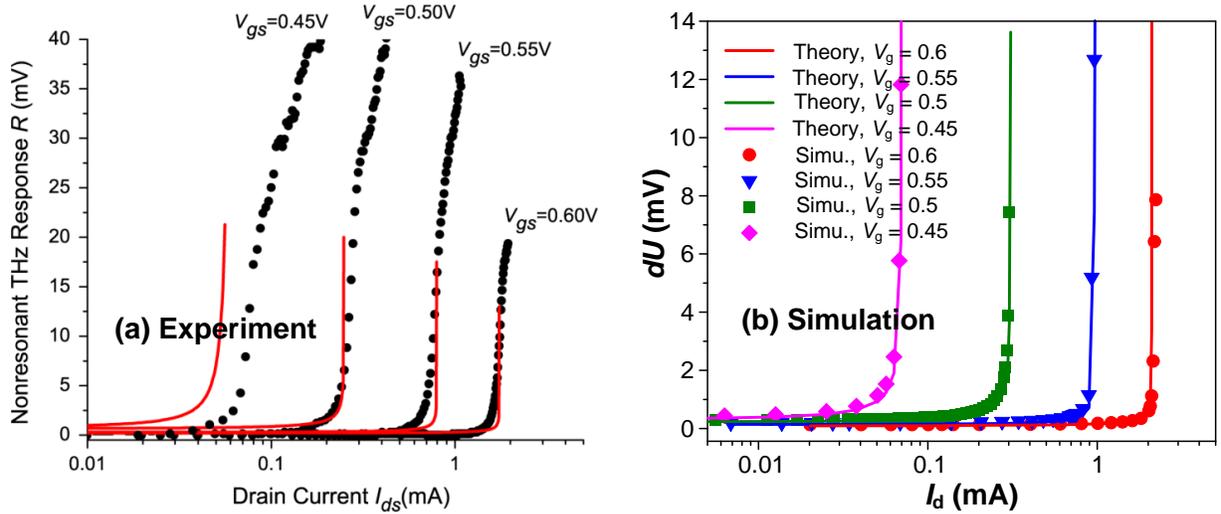

**Fig.1.** (a) Experimental data [35] and analytical results (black scatters) from THz detection of an InGaAs/GaAs HEMT with varying drain current under 4 different gate biases. (b) Simulation results of InGaAs/GaAs HEMT THz detection under the same operating conditions.

## 3. DC characteristics

Analogous to other high-frequency devices, the DC biasing condition strongly affects the AC performance of TeraFETs. The DC operation of a current-driven TeraFET can be described by [5]

$$\begin{cases} J_d = en_0 v_0 \\ \dfrac{\partial}{\partial x}\left(\dfrac{v_0^2}{2}+\dfrac{eU}{m}\right)+\gamma v_0 = 0 \end{cases} \tag{6}$$

where $n_0(x)$ and $v_0(x)$ are the DC steady-state carrier concentration and drift velocity, respectively. Eq. (6) needs to be solved along with the boundary condition (4). For $U_g \gg \eta V_t$, (4) becomes:

$$n_0(0) = \frac{C_g U_g}{e}, n_1(0) = \frac{C_g V_{am} \cos \omega t}{e}$$
$$n_1(L)v_0(L)+n_0(L)v_1(L)=0, J_d \approx en_0(L)v_0(L) \tag{7}$$

where $n_1(x)$ and $v_1(x)$ represent the carrier concentration and drift velocity induced by the small-signal AC voltage. If the convection term in (6) is ignored, we can get $J_d$ from (6) and (7) [5, 35]:

$$J_d = \frac{\mu C_g}{L}\left(\frac{U_g^2 - U_0^2(L)}{2}\right) = \frac{\mu C_g}{L}\left(U_g V_d - \frac{V_d^2}{2}\right), V_d \geq 0 \tag{8}$$

which is the classic Shockley model. Here $V_d = U_g - U_0(L)$ is the source-to-drain voltage difference. When $V_d = U_g = V_g - V_{th}$, we obtain the saturation current $J_{sat0} = \mu C U_g^2/2L$. Rearranging (8), the drain voltage swing $U_0(L)$ can be obtained as

$$U_0(L) = U_g \sqrt{1-\lambda}, \lambda < 1 \tag{9}$$

where $\lambda = J_d/J_{sat0}$. For an arbitrary position $x$, (9) can be generalized to $U_0(x) = U_g\sqrt{1-\lambda x/L}$ [5]. Therefore, the Shockley model predicts a square-root decrease of drain voltage swing as $J_d$ approaches the saturation criterion, and $U_0(L)$ or $V_d$ is irrelevant to the channel length. This expression is only valid for long-channel devices. For a short device, the hydrodynamic effects can lead to significant deviations from Eq. (9), as will be discussed later.

When the convection term is considered, (6) and (7) yields [55]:

$$\left(1-\frac{v_{cr}^3}{v_0^3}\right)\frac{\partial v_0}{\partial x}+\gamma = 0, \quad v_{cr}^3 = \frac{eJ_d}{m^* C_g} \tag{10}$$

where $v_{cr}$ is the so-called "choking velocity" [55]. Integrating Eq. (10) yields

$$\frac{v_{cr}^3}{2v_0^2(x)}+v_0(x) = \frac{v_{cr}^3}{2v_0^2(0)}+v_0(0)-x\gamma \tag{11}$$

Note that (10) and (11) are valid under $U_g \gg \eta V_t$ as the gradual channel approximation (GCA) and its simplified charge control $C_g U = en$ were used. In more general cases, one should replace GCA with UCCM, and then Eq. (10) changes to

$$\left[1 - \frac{1}{1-\exp(-J_d/\eta C_g V_t v_0)} \frac{v_{cr}^3}{v_0^3}\right] \frac{\partial v_0}{\partial x} + \gamma = 0, \quad v_{cr}^3 = \frac{eJ_d}{m^* C_g} \tag{12}$$

Eq. (12) can be solved numerically to obtain $v_0(x, J_d)$ and $n_0(x, J_d)$. Note that (12) is valid when the first term on the left-hand side is negative (since $n_0(x)$ or $U_0(x)$ should always drop monotonically from source to the drain under a positive $J_d$), i.e.

$$1 - \frac{1}{1-\exp(-J_d/\eta C_g V_t v_0)} \frac{v_{cr}^3}{v_0^3} < 0 \tag{13}$$

Recall that $J_d = en_0(L)v_0(L)$, $v_{cr}^3 = eJ_d/m^* C_g$, thus (13) yields

$$v_0(L) < \sqrt{\frac{\eta e V_t}{m^*}\left(1+\exp\left(-\frac{U_0(L)}{\eta V_t}\right)\right) \ln\left(1+\exp\left(\frac{U_0(L)}{\eta V_t}\right)\right)} = s(L) \tag{14}$$

where $s$ is the speed of the plasma wave [48]. Eq. (14) indicates that (12) is valid when the carrier velocity at the drain does not exceed the local plasma velocity, under which the plasmonic instability could occur. If instability occurs, the carrier density near the drain side oscillates at a given frequency, generating new harmonics and distorting the steady state. However, the electron-impurity scattering and viscosity effect suppress the growth rate of the instability [17, 44]. In our simulation, no significant instability phenomenon was observed.

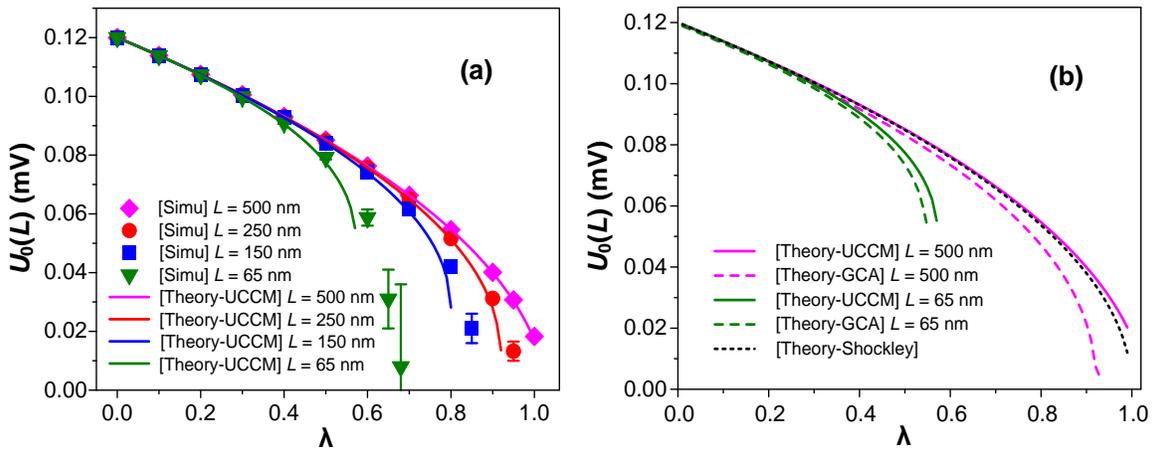

**Fig. 2.** Comparisons of (a) simulated response data and analytical theory curves (following Eq. (12)) and (b) analytical curves following (9), (11), and (12) under different channel lengths. $V_{gs} = 0.6$ V, $T = 300$ K.

Fig. 2(a) shows the comparison between the analytical curves following (12) and the numerical simulation results. As seen, good conformity between the theory and the simulation results is achieved. With the decrease in channel length, $U_0(L)$ drops significantly at $\lambda > 0.4$. Fig. 2(b) compares the analytical curves following (9), (11), and (12). For the Shockley model (9), the drain voltage does not change with $L$, thus it cannot properly describe $U_0(L, \lambda)$ in short channel devices ($L < L_0$, where $L_0 = s(\tau/\omega)^{0.5}$ [1, 56]). For (11), the use of GCA near the drain side (where $U_0(x)$ is small and GCA is invalid) results in a sharp drop of predicted $U_0(L)$. This leads to a larger mismatch with the simulation data compared to (12).

Since Eq. (12) is valid only when $v_0(L) < s(L)$ (i.e. the subsonic condition), the analytical curves terminate at $\lambda_{\text{ins}} = \lambda|_{v0(L) = s(L)}$. Beyond $\lambda_{\text{ins}}$, the solution of (12) is unphysical, but the simulation can still proceed. From the simulation data, we find that the actual saturation current $J_{\text{sat}}$ changes with $L$, as indicated in Fig. 2(a). Fig. 3 shows the ratio $J_{\text{sat}}/J_{\text{sat0}}$ as a function of $L$. As seen, $J_{\text{sat}}/J_{\text{sat0}}$ rises with the increase of $L$. When $L$ is sufficiently large, $J_{\text{sat}}$ can exceed $J_{\text{sat0}}$, but gradually saturates. Note that $J_{\text{sat0}} = \mu C U_g^2 / 2L = e\bar{n}_0 \bar{v}$, where $\bar{n}_0 = CU_g/2e = n_0(0)/2$, $\bar{v} = \mu \bar{E} = \mu U_g / L$. Since $J_{\text{sat0}}$ is based on GCA, which tends to underestimate the controlled electron density at low gate bias. At a high drain current, this effect becomes more pronounced. Therefore, we observe $J_{\text{sat}}/J_{\text{sat0}} > 1$ for long-channel devices.

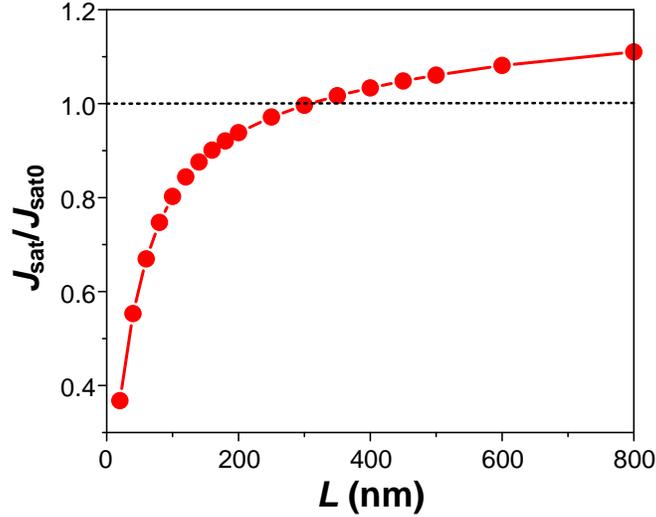

**Fig. 3.** $J_{\text{sat}}/J_{\text{sat0}}$ as a function of the channel length $L$. $J_{\text{sat0}} = \mu C U_g^2/2L$ is the saturation current predicted by the Shockley model.

We define $\lambda_{\text{sat}} = J_{\text{sat}}/J_{\text{sat0}}$ and $\lambda_1 = J_d/J_{\text{sat}} = \lambda/\lambda_{\text{sat}}$. Using these normalized parameters, the Shockley model (12) can be modified to

$$U_0(L) = U_g \left(1 - \lambda/\lambda_{sat}\right)^k = U_g \left(1 - \lambda_1\right)^k \tag{15}$$

Where $k$ is a fitting parameter. According to the simulation data in Fig. 2, $k \approx 0.32$ for $L = 65$, and $k \approx 0.5$ for $L > L_{cr}$. And $k$ roughly follows an interpolation relation with $L$ as $k(L) = -0.069 - 0.1 \cdot \ln(L-16.76)$. Fig. 4 shows $U_0(L)$ versus $\lambda/\lambda_{sat}$ under 4 channel lengths. A good match between simulation data points and (15) curves suggests that (12) can effectively describe the variation of DC drain voltage with $J_d$ under different channel lengths.

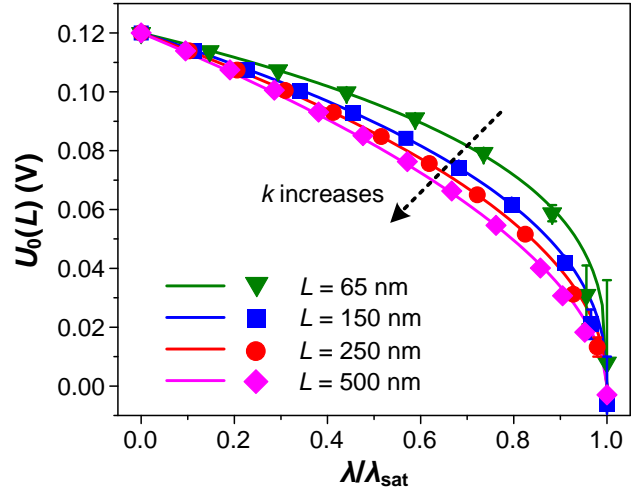

**Fig. 4.** DC drain voltage $U_0(L)$ as a function of $\lambda/\lambda_{sat}$ for 4 channel lengths. The scatters are simulation data while the solid curves are analytical curves following Eq. (15). For the 65 nm, 150 nm, 250 nm, and 500 nm channels, the values of the fitting parameter $k$ are 0.32, 0.42, 0.48, and 0.55, and the values of $J_{sat}$ are 0.68, 0.88, 0.97, and 1.05, respectively.

### 4. AC operation: long channel vs short channel

As was discussed in [5], both DC and AC performance of the TeraFET can be affected by the device channel length. The reason is related to the range of wave propagation inside the channel. When the plasma wave (or damped plasma oscillation) is generated near the source side and propagates toward the drain, the wave energy attenuates in the process due to local frictions. In a long enough channel, the traveling wave stops somewhere and cannot reach the drain. The critical propagation length $L_{cr}$ was define in [1, 56] as $L_{cr} = s\sqrt{\tau/\omega}$ for $\omega\tau \ll 1$ and $L_{cr} = s\tau$ for $\omega\tau \gg 1$. Now we suggest an interpolation equation accounting for the plasma wave (or damped plasmonic oscillation) propagation in all frequencies:

$$L_{cr} = s\tau\sqrt{(\omega\tau)^{-1} + 1} \tag{16}$$

which signifies the maximum distance that the plasmonic oscillation can travel before it is fully attenuated. Apparently, for $\omega\tau \gg 1$, $L_{cr} \approx s\tau$. For $\omega\tau \ll 1$, $L_{cr} \approx s(\tau/\omega)^{0.5}$. Using $L_{cr}$ as a criterion, the TeraFET detectors can be categorized into two groups: the resonant detector ($L \ll L_{cr}$), and the broadband detector ($L \gg L_{cr}$). For current-driven TeraFETs, distinctive behaviors were predicted in these two types of detectors [5]. Later we'll discuss those behaviors in more detail. Fig. 5 presents $L_{cr}$ as a function of frequency for the subject InGaAs/GaAs TeraFETs and compares $L_{cr}$ with 4 different channel lengths ranging from 65 nm to 500 nm. For a long channel sample, such as the one with $L = 500$ nm, $L_{cr}$ drops below $L$ at a very low frequency ($f < 0.1$ THz). With the increase of $L$, the critical frequency for $L = L_{cr}$ rises rapidly. At $L = 65$ nm, $L < L_{cr}$ always holds. As will be shown later, the relative size between $L$ and $L_{cr}$ strongly affects the THz detection of current-driven TeraFETs.

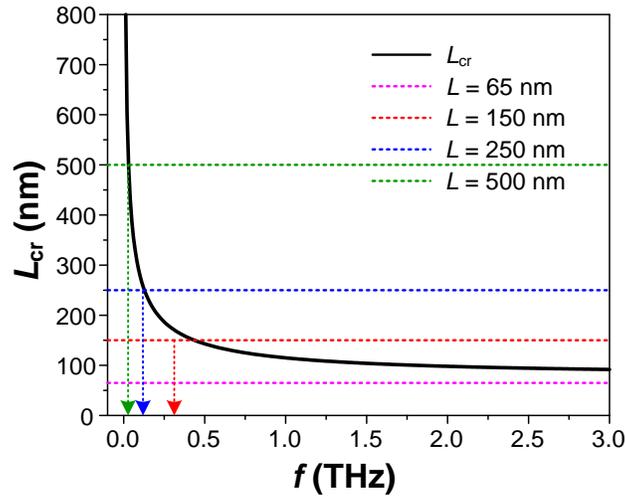

**Fig. 5.** The critical channel length $L_{cr}$ as a function of $f$. The dash lines show the level of 4 different channel lengths

Now we evaluate the THz detection performance of current-driven InGaAs/GaAs TeraFETs with different channel lengths. Fig. 6 presents the simulation results of $dU$ versus $f$ under different drain current biases with $L = 65\sim500$ nm. For better comparison, Fig. 7 shows $dU$ versus $\lambda$ at different frequencies. When $L = 500$ nm, $L \gg L_{cr}$ holds except for very low frequencies (e.g. $f < 0.1$ THz). In this case, $dU$ rises sharply with the increase of frequency in the low-frequency band, and gradually stabilizes, as shown in Fig. 7(a). The negative response can be observed only within $f < 0.1$ THz. According to [5], the negative response is a signature of short-channel or low-frequency detection. Beyond 0.1 THz, the response rises sharply with increasing $\lambda$, as shown in Fig. 7(a). While $\lambda$ rises from 0 to $\lambda_{sat}$, $dU$ experiences an order of magnitude elevation. The above results signify the THz detection characteristics of a typical long-channel TeraFET under drain current bias, which agree with those in [5] and [54].

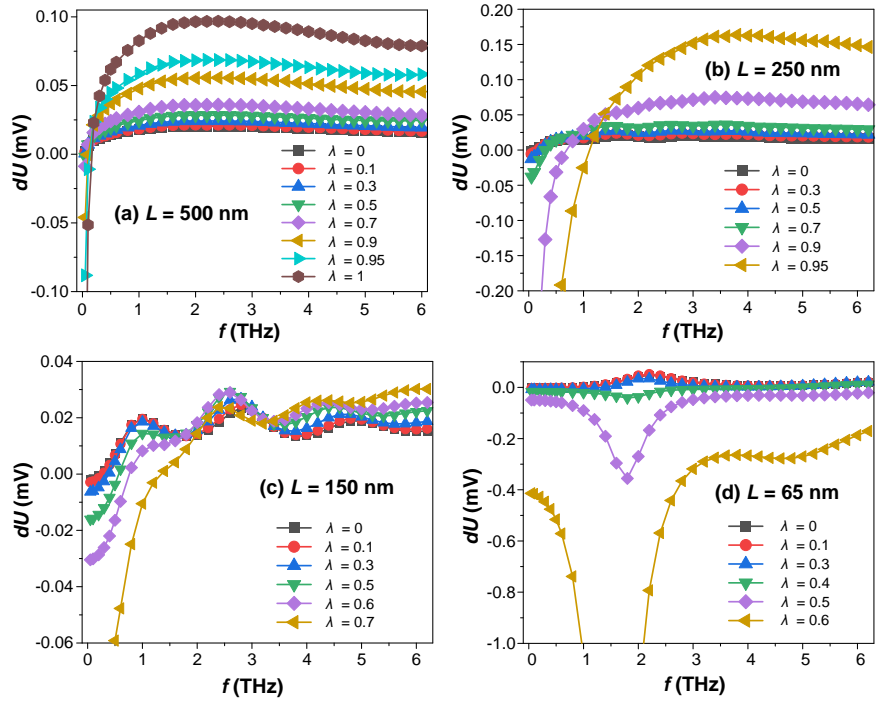

**Fig. 6.** Simulation results of $dU$ as a function of $f$ at $\lambda = 0\sim1$ with 4 different channel lengths. Here $V_{gs} = 0.6$ V, $T = 300$ K. Other unspecified parameters follow Table I.

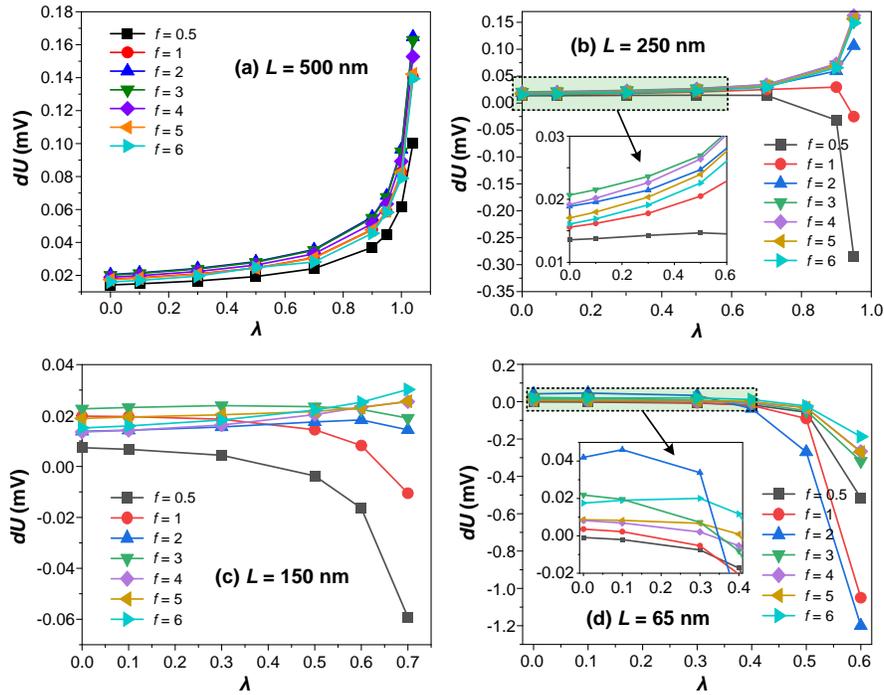

**Fig. 7.** Simulation results of $dU$ as a function of $\lambda$ at $f = 0.5\sim6$ THz with 4 different channel lengths.

As $L$ decreases, the frequency range in which $L < L_{cr}$ expands, so does the band of negative response, as seen in Fig. 6(b) and Fig. 6(c). In the negative response band, $dU$ drops drastically with the increase of $\lambda$, as shown in Fig. 7(b), 7(c).

We now review some insights into the negative response phenomenon. Under the non-resonant condition ($\omega\tau < 1$), the DC voltage response $dU = U_0(L) - U_0(L)|_{Vam=0}$ can be analyzed by substituting the continuity equation (1) into (2) while ignoring the convection term and time-dependent term. Now (1) and (2) simplify to [3]:

$$v = -\frac{e\tau}{m^*}\frac{\partial U}{\partial x} = -\mu\frac{\partial U}{\partial x},$$
$$\frac{\partial U}{\partial t} - \frac{\mu}{2}\frac{\partial^2 U^2}{\partial x^2} = 0 \tag{17}$$

Note that GCA is used here ($C_g U = en$). One can search the solution of $U$ in the form of $U(x) = U_0(x) + \frac{1}{2}U_1(x)\exp(-i\omega\tau) + \frac{1}{2}U_1^*(x)\exp(i\omega\tau)$, where $U_1(x)$ represents the first-order voltage oscillation induced by the THz signal. Solving (17) along with the boundary condition (4) yields [5]:

$$U_0^2(x) + \frac{|U_1(L)|^2}{2} = U_g^2 + \frac{V_{am}^2}{2} - \frac{2J_d}{\mu C_g}x \tag{18}$$

By setting $V_{am} = 0$ or $V_{am} \neq 0$, one can get $U_0(L)$ and $U_0(L)|_{Vam=0}$, and thereby $dU$:

$$dU = U_0(L) - U_0(L)|_{V_{am}=0} \approx \frac{V_{am}^2 - |U_1(L)|^2}{4U_g\sqrt{1-\lambda}} \tag{19}$$

In a long-channel device, the plasmonic oscillation cannot reach the drain, hence $U_1(L) = 0$. Consequently, the response under $L \gg L_{cr}$ is given by (after normalized with $\lambda_{sat}$)

$$dU = \frac{V_{am}^2}{4U_g}\frac{1}{\sqrt{1-\lambda/\lambda_{sat}}} \tag{20}$$

In this case, the response is always positive and is inversely proportional to the drain voltage swing $U_0(L) \approx U_g(1-\lambda/\lambda_{sat})^{0.5}$. For $L < L_{cr}$, the plasmonic oscillation can reach the drain and $U_1(L) \neq 0$. According to [5], $U_1(L)$ can be solved using (17) and related boundary conditions:

$$dU \approx \frac{V_{am}^2}{4U_g}\frac{1}{(1-\lambda/\lambda_{sat})^{1.5}}(-\lambda/\lambda_{sat} + P(\lambda/\lambda_{sat})\varepsilon_0^2),$$
$$P(\lambda/\lambda_{sat}) = k_p\frac{4\sqrt{1-\lambda/\lambda_{sat}} + 6 - \lambda/\lambda_{sat}}{(1+\sqrt{1-\lambda/\lambda_{sat}})^4} \tag{21}$$

where $\varepsilon_0 = (L/L_{cr})^2 = \omega L^2/S^2\tau = \omega/\omega_0$, $\omega_0 = S^2\tau/L^2$ is a critical angular frequency, and $k_p$ is a fitting constant. The variable $P$ is a function of $\lambda/\lambda_{sat}$, which is similar to the weighting factor in the MOSFET charge control model [57]. As $\lambda$ rises, $dU$ turns from positive to negative. The origin of this evolution is related to the competition between current-induced field effects and damped plasmonic contribution [5]. At zero drain current, (21) reduces to $dU = (V_{am}^2/4U_g)\cdot P(0)\varepsilon_0^2$, which is always positive and contributed solely by the damped plasmonic oscillations. Under a finite drain-to-source current, the field-induced term $(V_{am}^2/4U_g)(-\lambda/\lambda_{sat})/(1-\lambda/\lambda_{sat})^{1.5}$, which is always negative, starts contributing to $dU$, and the amplitude of this term grows with rising $\lambda$. With a sufficiently large $\lambda$, $dU$ turns negative. From (19), we can see that when $dU < 0$, $|U_1(L)|^2 > V_{am}^2 = |U_1(0)|^2$. This suggests that the small-signal AC voltage is amplified throughout the channel when the drain current is sufficiently high. Therefore, the device now serves as a power amplifier for the THz signal.

The amplification of THz signal power is verified by checking the simulated time-domain voltage swing at the drain as compared to that of the source, as shown in Fig. 8. Since the first-order term predominates, we can assume $U_1(L) \approx U(L) - U_0(L)$. As seen in Fig. 8, the amplitude of drain voltage swing exceeds the one at the source when $L$ = 80 nm and 200 nm, indicating an AC voltage/power gain. The absorbed power comes from the drain bias.

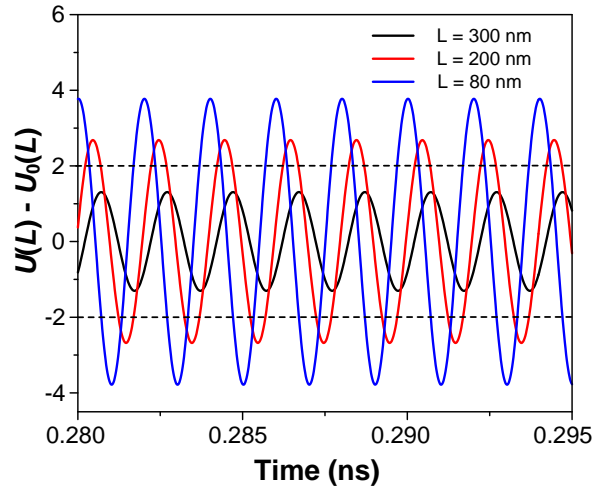

**Fig. 8.** Simulated temporal waveforms of $U(L) - U_0(L)$ under $\lambda_{sat}/\lambda = 0.7$, $f = 0.5$ THz (non-resonant), $V_{am}$ = 2 mV, $L$ = 80-300 nm. Other unspecified parameters are the same as those in Fig. 6. The black dashed lines show the boundaries of voltage swing at the source (i.e. $\pm V_{am}$).

To analyze this power gain, we reformulate (19) and obtain $|U_1(L)|^2/2 - V_{am}^2/2 \approx -2dU\cdot U_0(L)$. Obviously, $|U_1(L)|^2/2$ and $V_{am}^2/2$ are proportional to the powers of AC voltage at the drain and

source side, respectively. The right-hand side term $-2dU \cdot U_0(L)$ signifies the AC power gain/loss throughout the device channel. A negative $dU$, by definition, suggests a positive radiation-induced drain-to-source DC voltage. As a result, the AC signal at the drain has a higher average power as compared to that at the source. In addition, we also notice a phase change in $U(L)$ when $L$ varies. As $L$ increases, the phase of $U(L)$ delays. This phase shift should be related to the traveling time of plasmonic oscillations from the source to the drain, or the source-to-drain delay time.

Fig. 9 shows the channel length dependence of $dU$ under different drain biases. As seen, for $\lambda_{sat}/\lambda > 0$ the response becomes negative at the low $L$ region, and a minimum of $dU$ is observed in each curve. Recall that in the limit of $L \to 0$, one gets $dU \to 0$ since $|U_1(L)| \approx V_{am}$. As $L$ increases, $|U_1(L)|$ gets amplified by the current-induced field and becomes larger than $V_{am}$, thus $|dU|$ starts to increase. When $L$ approaches $L_{cr}$, the attenuation of $|U_1(x)|$ by scatterings gradually overshadows the power gain from the DC field, and then both $|U_1(L)|$ and $|dU|$ decrease. Finally, $|U_1(L)|$ drops below $V_{am}$ and $dU$ turns from negative to positive. When $L$ exceeds $L_{cr}$, $|U_1(L)|$ becomes very small, and the rise of $dU$ with increasing $\lambda_{sat}/\lambda$ is contributed solely by the drop of $U_0(L)$.

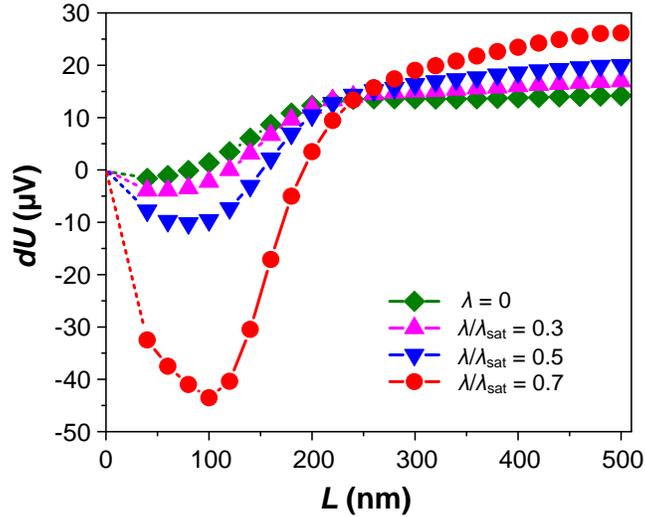

**Fig. 9.** $dU$ as a function of $L$ under 4 different drain biases. Here $f = 0.5$ THz (non-resonant), $V_{am} = 2$ mV. Other unspecified parameters are the same as those in Fig. 8.

The mechanism behind the aforementioned amplification of $U_1$ is related to the non-uniform electron density distribution generated by the DC field. A similar effect has been discussed in our previous paper [23], where the non-uniform carrier distribution created a density gradient that altered the transport properties and affected the device response. This THz amplification mechanism is different from those in the conventional HEMT THz amplifiers [58-61] or the resonant tunnel diode (RTD) gated plasmonic amplifiers [62, 63]. In a conventional HEMT THz amplifier, the intrinsic capacitances, resistances, electron transit time, etc. are optimized to achieve a higher transducer gain in the THz or sub-THz range [59, 60]. For the RTD-assisted plasmonic amplifiers, the THz amplification is achieved by the coupling of the RTD gate and the plasma wave channel [62, 63].

To qualitatively analyze the DC field effect, we first evaluate the field distribution $E_0(x)$. Using $U_0(x)$ under the Shockley model, the current-induced DC field $E_0(x)$ and $E_0(L)$ can be obtained via

$$E_0(L) = -\frac{\partial V_{ch}(x)}{\partial x}\bigg|_{x=L} = \frac{\partial U_0(x)}{\partial x}\bigg|_{x=L} = -\frac{U_g}{2\sqrt{1-(\lambda/\lambda_{sat})(x/L)}}\frac{\lambda}{\lambda_{sat}}\frac{1}{L}\bigg|_{x=L}$$
$$= E_0(x)\big|_{x=L} = -\frac{U_g}{2\sqrt{1-\lambda/\lambda_{sat}}}\frac{\lambda}{\lambda_{sat}}\frac{1}{L} \tag{22}$$

The negative sign suggests that the direction of $E_0(x)$ is drain-to-source. We can see that $E_0(x)$ is highly non-uniform along the channel, and $E_0(x)$ increases sharply near the drain side. Comparing (22) with the expression of $dU|_{\varepsilon_0=0}$, we find that

$$dU\big|_{\varepsilon_0=0} = \frac{V_{am}^2}{2U_g^2}\frac{E_0(L)\cdot L}{1-\frac{\lambda}{\lambda_{sat}}} = \frac{V_{am}^2}{4U_0(L)}\cdot\frac{E_0(L)\cdot 2L}{U_0(L)} \tag{23}$$

Note that $U_0(L) - U_0(0) = U_0(L) - U_g = \int_0^L E_0(x)dx$, therefore,

$$dU\big|_{\varepsilon_0=0} = \frac{V_{am}^2}{4U_0(L)}\cdot\frac{2E_0(L)\cdot L}{U_g + \int_0^L E_0(x)dx} = \frac{V_{am}^2}{4U_0(L)}\cdot\frac{2E_0(L)}{U_g/L + \overline{E_0}} \tag{24}$$

where $\overline{E_0} = \int_0^L E_0(x)dx/L$ is the average lateral DC field along the channel. Inspecting Eq. (24), we see that the first multiplier on the r.h.s. is the response voltage for long-channel devices. The second multiplier on the r.h.s. is related to the ratio of $E_0(L)$ over $\overline{E_0}$, which signifies the nonuniformity of the electric field in the channel. Therefore, the amplitude of $dU$ is directly controlled by the nonuniformity of $E_0(x)$. This result agrees with our observations in [23].

The above equations are based on the Shockley model. Using (15), the drain field becomes

$$E_0(L) = -kU_g\left(1-\frac{\lambda}{\lambda_{sat}}\right)^{k-1}\cdot\frac{\lambda}{\lambda_{sat}}\frac{1}{L} \tag{25}$$

It is obvious that when $k = 0.5$, (25) reduces to (22). Substituting (25) into (23), we get

$$dU\Big|_{\varepsilon_0=0} = \frac{V_{am}^2}{4U_0(L)} \cdot \frac{E_0(L) \cdot 2L}{U_0(L)} = -\frac{V_{am}^2}{4U_g} \frac{2k}{\left(1-\frac{\lambda}{\lambda_{sat}}\right)^{k+1}} \cdot \frac{\lambda}{\lambda_{sat}} \tag{26}$$

Based on (26), we generalize (21) into

$$dU \approx \frac{V_{am}^2}{4U_g} \frac{2k}{(1-\lambda/\lambda_{sat})^{k+1}} (-\lambda/\lambda_{sat} + P(\lambda/\lambda_{sat})\varepsilon_0^2) \tag{27}$$

Compared to (21), (27) offers better compatibility for detections under different operating conditions.

Note that (27) is valid only when $\omega\tau \ll 1$. For the resonant detection, as shown in Fig. 6(d), the response curves preserve the resonant peaks, but with negative amplitudes. Therefore, we can approximate $dU$ by a semi-empirical expression as follows:

$$dU \approx \frac{V_{am}^2}{4U_g} \frac{2k\left(-\lambda/\lambda_{sat} + P(\lambda/\lambda_{sat})\varepsilon_0^2\right)}{(1-\lambda/\lambda_{sat})^{k+1}} f(\omega) \tag{28}$$

Where $f(\omega)$ is the frequency-dependent function of resonant detection [1, 48]. Eq. (28) indicates that with the onset of plasmonic resonance, the effect of the current-induced electric field is further enhanced, thus the resulting negative response is amplified by the factor $f(\omega)$, which could exceed 100 or even higher near the fundamental resonant peak.

With the viscosity effect included, $f(\omega)$ is given by [48]

$$f(\omega) = 1 + \beta - \frac{\beta\left(\cos^2 k_1 L - \frac{\tau}{\tau_1}\sinh^2 k_1 L\right) + \cos 2k_1 L}{\sinh^2 k_1 L + \cos^2 k_1 L} \tag{29}$$

where $\tau_1 = \upsilon/s^2$ is the viscosity time constant, $\beta = (2\omega^2/s^2)/(k_1^2 + k_2^2)$, and $k_1$ and $k_2$ are wave vectors of the plasma wave obtained from the dispersion relation [48]:

$$k_1, k_2 = \frac{\omega}{s\sqrt{1+(\omega\tau_1)^2}} \frac{1}{\sqrt{2}} \sqrt{\sqrt{\left(1-\frac{\tau}{\tau_1}\right)^2 + \left(\frac{1}{\omega\tau}+\omega\tau_1\right)^2} \pm \left(1-\frac{\tau}{\tau_1}\right)} \tag{30}$$

Another issue worth addressing is the shift of resonant peaks with varying $J_d$, as shown in Fig. 6(d). This phenomenon is due to the decrease of $s(x)$ with increasing $J_d$, since $s(x)$ is a function of $n_0(x)$ (see Eq. (3)). Note that the position of resonant peaks is given by $f_n = ns/4L$, where $n = 1, 3, 5, \ldots$ is the order of odd harmonics. Therefore, $f_n$ drops with the decrease of $s$. Since $s(x)$ is now nonuniform, it is difficult to analytically quantify the exact $s(x)$. Here we assume that the dc response is related to the average of $s$ over the channel, $\bar{s}$. Based on our simulation data, we can approximate by $\bar{s} = s_0(1+(1-\lambda/\lambda_{sat})^b)^{0.5}$, where $s_0$ is the plasma velocity under zero drain current, $b \approx 0.2$ is a fitting parameter. Then we replace $s$ in (30) with $\bar{s}$.

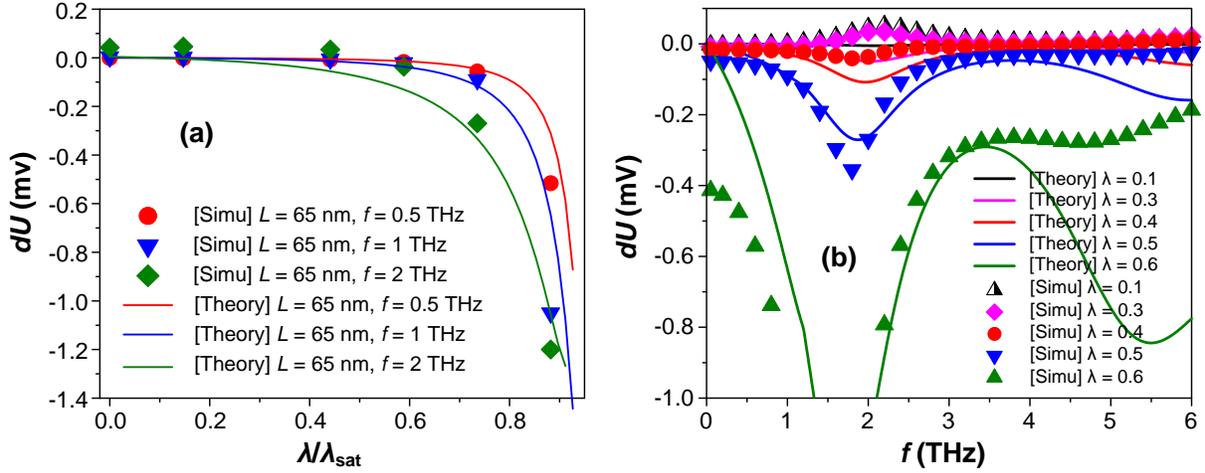

**Fig. 10.** (a) $dU$ as a function of $\lambda/\lambda_{sat}$ under $L = 65$ nm, $f = 0.5\sim2$ THz. (b) $dU$ as a function of $f$ under $\lambda = 0.3\sim0.6$ ($\lambda/\lambda_{sat} = 0.44\sim0.88$). The theory curves in (a) and (b) follow (19).

Fig. 10(a) shows the comparison of analytical response curves following (28) and the simulation data as functions of $\lambda/\lambda_{sat}$ in a relatively low-frequency range ($f \leq 2$ THz). A fair quantitative agreement between the simulation and analytical results is reached. This shows that (28) is valid in the low-frequency range.

Fig. 10(b) shows the frequency-dependent of $dU$ obtained from (28) and the simulation data. In general, the analytical curves can qualitatively predict the variation trend of $dU$. For $\lambda = 0.5$ and 0.6, a relatively good quantitative agreement is reached at $f = 0.5\sim3.5$ THz. In the vicinity of the second harmonic peak, the simulated response is much higher compared to the analytical value, indicating a suppression of higher-order harmonics that is unconsidered in (28). Besides, at the low-frequency band the simulated response is significantly larger than the prediction of (28). Analysis of the subtle mismatch between (28) and the simulation is beyond the scope of this paper and will be studied in our future works.

## 5. Current-driven TeraFET spectrometer

As demonstrated in the last section, in a short-channel TeraFET, the transition between positive and negative response voltages is controlled by the THz frequency and drain current. Therefore, we can relate the polarity transition frequency $f_{trand}$ (i.e. $f_{tr} = f|_{dU=0}$) with the drain current ratio $\lambda/\lambda_{sat}$. According to Eq. (27) or Eq. (28), when $dU = 0$, we get

$$\omega_{tr} = \sqrt{\frac{\lambda_{cr}}{\lambda_{sat}} \frac{c}{P(\lambda_{cr}/\lambda_{sat})}} \omega_0, \quad \lambda_{cr} = \lambda|_{dU=0} \tag{31}$$

where $\omega_{tr} = 2\pi f_{tr}$ is the angular transition frequency, and $c$ is a fitting parameter, which is obtainable via calibration. Therefore, in theory, we can get the frequency of an unknown THz signal by the following simple steps: 1) calibrate the system parameter $c$ using standard frequency signals; 2) detect the target signal, adjust the bias current $I_d$ until $dU = 0$; 3) obtain $\lambda|_{dU=0}$ as the $\lambda_{cr}$; (4) calculate $\omega_{tr}$ using (31) and take $\omega_{pred} = \omega_{tr}$, where $\omega_{pred}$ is the predicted frequency of the target signal. Convert the angular frequency $\omega_{pred}$ to ordinary frequency $f_{pred}$ if necessary.

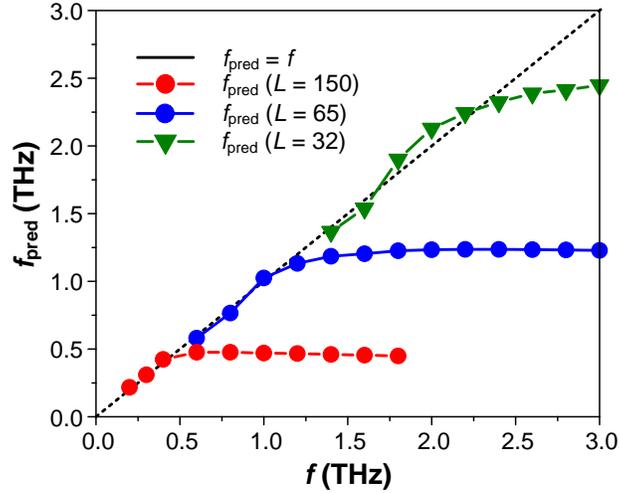

**Fig. 11.** The predicted frequency using (31) as a function of $f$. The short-dashed line shows $f_{pred} = f$ line.

Fig. 11 shows the calculated $f_{pred}$ following the above steps as a function of setup frequency $f$ under $L = 32$ nm, 65 nm, and 150 nm. The black dashed line illustrates the ideal $f_{pred} = f$ line. As seen, in each device, the calculated $f_{pred}$ values are in line with $f$ only in the low-frequency region, in which $\omega \ll \omega_0$ (or, equivalently, $L \ll L_{cr}$). As the frequency approaches and rises beyond $\omega_0$, $f_{pred}$ deviates from the actual frequency and saturates in the end. Therefore, the use of the aforementioned spectrometer is limited to the frequency range of $\omega \ll \omega_0$.

Despite having a limited applicable frequency range, the $I_d$-driven TeraFET spectrometer still has its advantages compared to other TeraFET-based spectrometers. One of the most significant advantages is the ease of implementation. Previously, the realization of TeraFET-based spectrometers relied on spatial nonuniformity [45] or boundary asymmetry [28, 29]. For example, in [45], a "defect" 2DEG channel was created using electron density steps. With this defection, a photovoltage oscillation was induced under an incident sub-THz radiation, and the oscillation frequency was a function of the sub-THz signal frequency. In [28] and [29], the proposed TeraFET spectrometers required the phase asymmetry in the THz-induced voltage at source and drain, which was realizable by using 2 different antennas. Neither the electron density steps nor the phase-separated terminal antennas are as convenient as the DC drain current in our spectrometer in terms of tunability. In addition, the calibration of previous spectrometers can be significantly more challenging, since the characterizations of electron density distribution and the phase difference are more difficult compared to the drain current adjustments. With those advantages, the $I_d$-driven TeraFET has the potential to become an effective spectrometer in the sub-THz band.

## 6. Conclusion

In this work, a verified 1D hydrodynamic model was used to study the characteristics of an InGaAs/GaAs TeraFET under DC drain current bias. The major conclusions are listed below:

1. In the DC operation, the drain voltage dropped as the channel length shortens, thus deviating from the Shockley model. A new analytical equation describing the DC current-voltage relation in TeraFETs was derived, which agreed better with the simulation values compared to the Shockley model. We also developed a new semi-empirical expression to describe the DC drain voltage as functions of channel length and bias current.

2. The continuous-wave THz detection performance of DC-current-biased TeraFETs showed a strong channel length dependence. For a sufficiently short device or at a sufficiently low frequency, the source-to-drain voltage response $dU$ could turn negative. The negative $dU$ results from the non-uniform field distribution due to the DC current. In the meantime, the radiation-induced small-signal voltage is also amplified throughout the channel, and the device now can be used as a THz amplifier. Under the resonant condition, the response curves preserved the resonant peaks, but with negative amplitudes. A semi-empirical expression was proposed to qualitatively describe the variation of $dU$ in this condition.

3. The short-channel TeraFET has the potential to be used as the THz spectrometer driven by the drain current.